# Decay pion spectroscopy: a new approach

A. Margaryan, R. Ajvazyan, N. Grigoryan, V. Kakoyan, V. Khacatryan,
H. Vardanyan, S. Zhamkochyan
A.I. Alikhanyan National Science Laboratory, Yerevan, Armenia

P. Achenbach, J. Pochodzalla
Institut fur Kernphysik, Johannes Gutenberg-Universitat, 55099 Mainz, Germany

S.N.Nakamura, S.Nagao*, Y.Toyama*
Graduate School of Science, Tohoku University, Sendai 980-8578, Japan
* Gradute Program on Physics for the Universe (GP-PU), Tohoku University

J.R.M. Annand, K. Livingston, R. Montgomery
School of Physics & Astronomy, University of Glasgow, G128QQ Scotland, UK

Keywords: light hypernuclei, decay pion spectroscopy, recoil distance technique, magnetic spectrometer, low pressure MWPCs

**Abstract**

We propose a new experiment for decay pion spectroscopy of light hypernuclei at electron- and proton-beam facilities, using the recoil distance technique for separation of produced hypernuclei and a magnetic spectrometer for precise measurement of the decay pion momentum. Low-pressure MWPCs are advocated for low-energy recoil detection as they provide position and time information and are highly insensitive to gamma-ray and electron background. The position and timing characteristics of such a recoil detector were studied using ~5 MeV α-particles. By using the present proposed approach the rate of the detected hypernuclei can be increased by one-to-two orders of magnitude compared to a recent experiment carried out at the MAMI electron accelerator in Mainz. The possibility of realizing a high luminosity decay pion spectroscopy experiment with proton beams is also discussed.

## 1. INTRODUCTION

The binding energy of the Λ particle in the hypernuclear ground state gives one of the basic pieces of information on the Λ-nucleus interaction. This binding energy is defined as:

$$B_{\Lambda}(g.s.) = M_{core} + M_{\Lambda} - M_{HY}.$$

where $M_{core}$ is the mass of the nucleus that is left in the ground state after the $\Lambda$ particle is removed, $M_{HY}$ is the mass of the initial hypernucleus and $M_{\Lambda}$ is the mass of the $\Lambda$. The binding energies $B_{\Lambda}$ have been measured in emulsion-based experiments for a wide range of light ($3 \leq A \leq 15$) hypernuclei [1-3], exclusively from kinematic analysis of their weak $\boldsymbol{\pi}^{-}$ -mesonic decays. The binding energies of light hypernuclei provide the most valuable experimental information to constrain various models of the Y-N interaction. Table 1, where the numbers are taken from reference [4], lists the Λ separation energies obtained from *ab initio* theoretical calculations using different Y-N interactions, along with the existing experimental results. In addition to the quoted statistical errors, the experiments also have systematic errors of about 0.04 MeV. Precise experimental measurements of the binding energies of light hypernuclei can discriminate between various models of Y-N interactions. In particular, accurate measurements of the separation energies of light Λ-hypernuclei are a unique source of information on charge symmetry breaking in the Λ –N interaction in Λ-hypernuclei [5, 6].

Table 1. Λ separation energies, $B_{\Lambda}$ given in units of MeV, of A = 3-5 Λ hypernuclei for different Y-N interactions.

| Y-N | $B_{\Lambda}({}^{3}_{\Lambda}H)$ | $B_{\Lambda}({}^{4}_{\Lambda}H)$ | $B_{\Lambda}({}^{4}_{\Lambda}H^{*})$ | $B_{\Lambda}({}^{4}_{\Lambda}He)$ | $B_{\Lambda}({}^{4}_{\Lambda}He^{*})$ | $B_{\Lambda}({}^{5}_{\Lambda}He)$ |
|---|---|---|---|---|---|---|
| **SC97d(S)** | 0.01 | 1.67 | 1.20 | 1.62 | 1.17 | 3.17 |
| **SC97e(S)** | 0.10 | 2.06 | 0.92 | 2.02 | 0.90 | 2.75 |
| **SC97f(S)** | 0.18 | 2.16 | 0.63 | 2.11 | 0.62 | 2.10 |
| **SC89(S)** | 0.37 | 2.55 | Unbound | 2.47 | Unbound | 0.35 |
| **Experiment** | 0.13±0.05 | 2.04±0.04 | 1.00±0.04 | 2.39±0.03 | 1.24±0.04 | 3.12±0.02 |

In 2007 the use of magnetic spectrometers to measure precisely the momenta of pions from weak, two-body decays of electroproduced hyperfragments was proposed for Jefferson Lab [7, 8]. A similar experimental program was started at MAMI in Mainz [9], where the first high resolution spectroscopy of pions from decays of strange systems was performed after electro-disintegration of a ${}^{9}Be$ target [10-12]. About $10^{3}$ weak pionic decays of hyperfragments and hyperons were observed in a two-week period (~150 h). The pion momentum distribution shows a monochromatic peak at p ≈ 133 MeV/c, corresponding to the unique signature for the two-body decay of hyper hydrogen ${}^{4}_{\Lambda}H \rightarrow {}^{4}He + \pi^{-}$, where the ${}^{4}_{\Lambda}H$ stopped inside the target. Its binding energy was determined to be $B_\Lambda = 2.12 \pm 0.01$ (stat.) ± 0.09 (syst.) MeV with respect to the $^3H + \Lambda$ mass.

In the 2012 and 2014 experiments the MAMI beam, with an energy of 1.5 GeV and an current of 20-50 μA, was incident on a beryllium foil of either 125 mm or 250 mm thickness. The $^9Be$ target was tilted by 54° with respect to the beam direction to minimize the energy straggling of negative pions leaving the target in the direction of the spectrometers. A significantly thicker target with the same arrangement would have caused too large a spread of the pion momentum. Also, an increase of the luminosity could not be achieved by an increase of the beam current, due to high counting rates in the forward angle kaon spectrometer. Even a factor of two increase of the beam intensity would have raised the count rate caused by background coincidences beyond an acceptable level.

Recently a new experiment was proposed at MAMI for the study of the binding energy of the Λ in the lightest hypernucleus: the hypertriton, which can reveal important details of the strong nucleon-hyperon interaction. In this experiment a new design of lithium target was considered to increase the rate of the detected hypernuclei by about an order of magnitude [13].

We propose a new experiment to perform decay pion spectroscopy of light hypernuclei , using the recoil distance technique for separation of produced hypernuclei and a magnetic spectrometer for precise measurement of the decay pion momentum. Low-pressure MWPCs are proposed for low-energy recoil detection since they provide precise position and time information, and are highly insensitive to electrons and gamma rays. Possible experimental studies at the MAMI electron-scattering facility and the JPARC proton beam facility are discussed.

The decay pion experiments at MAMI are described in Sec. 2. The proposed new experiment at MAMI is presented in Sec. 3. In Sec. 4 the recoil detector is described and measurements of the detector's response to ~5 MeV α-particles are presented. Sec. 5 is devoted to the MC simulations. In Sec. 6 possible background sources are discussed. The expected yields at electron and proton beam facilities are considered in Sec.7. Practical issues and future possibilities are discussed in Sec. 8.

## 2. THE HYPERFRAGMENT ELECTROPRODUCTION EXPERIMENT AT MAMI-C

This experiment was carried out by the A1 Collaboration at the spectrometer facility at the Mainz Microtron MAMI [14]. A 1.508 GeV electron beam with a current of 20 μA was incident on a 125 μm thick $^{9}Be$ target foil angled at 54 degrees with respect to the beam direction. Pions were detected by two high-resolution spectrometers (SpekA and SpekC), each having a quadrupole-sextupole-dipole-dipole(QSDD) configuration (FIG. 1, left).

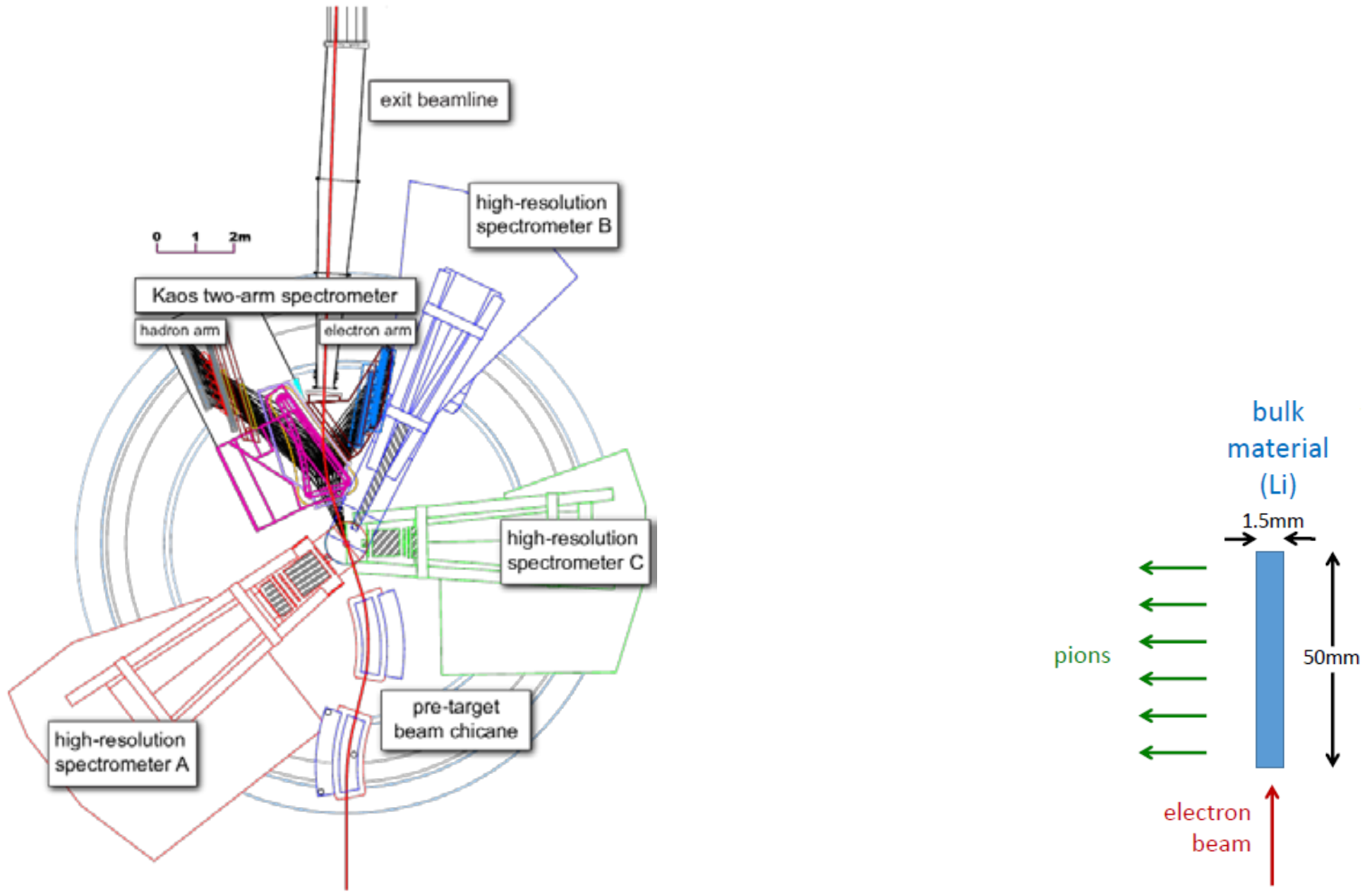

FIG. 1. Left: floor plan of the electron beam-line and magnetic spectrometers in the experimental hall at the Mainz Microtron MAMI. Right: schematic of the lithium target, which is proposed for new experiments at MAMI.

The spectrometers achieve a relative momentum resolution of $\delta p/p \approx 10^{-4}$ and were operated at central momenta of 115 (SpekA) and 125 (SpekC) MeV/c with momentum acceptances of $\Delta p/p = 20$ % (SpekA) and 25 % (SpekC) with a solid angle acceptance 20 - 30 msr. The tagging of kaons was performed by the Kaos spectrometer, positioned at zero degrees with respect to the electron beam direction. The central momentum was 924 MeV/c, covering a momentum range of $\Delta p/p = 50$ % with a solid angle acceptance of $\Omega_K = 16$ msr.

The Fig. 2 shows the pion momentum distribution in SpekC for the events selected within the $K^+ - \pi^-$ coincidence time gate. The distribution above the accidental background is attributed to mesonic weak decay (MWD) events. About $10^3$ pionic weak decays of hyperfragments and quasi-free produced hyperons were observed within the entire momentum acceptance of SpekC. The pion momentum distribution shows a monochromatic peak at ~133 MeV/c, corresponding to the unique signature of about 50 two-body decays of ${}^{4}_{\Lambda}H \rightarrow {}^{4}He + \pi^{-}$. From this the ${}^{4}_{\Lambda}H$ binding energy was determined.

No localized excess of counts over expectations of background from quasi-free hyperon decays were found in the pion momentum distribution of SpekA. This was set to detect decay pions in the momentum range 103-127 MeV/c, which is outside the region of interest for ${}^{4}_{\Lambda}H$.

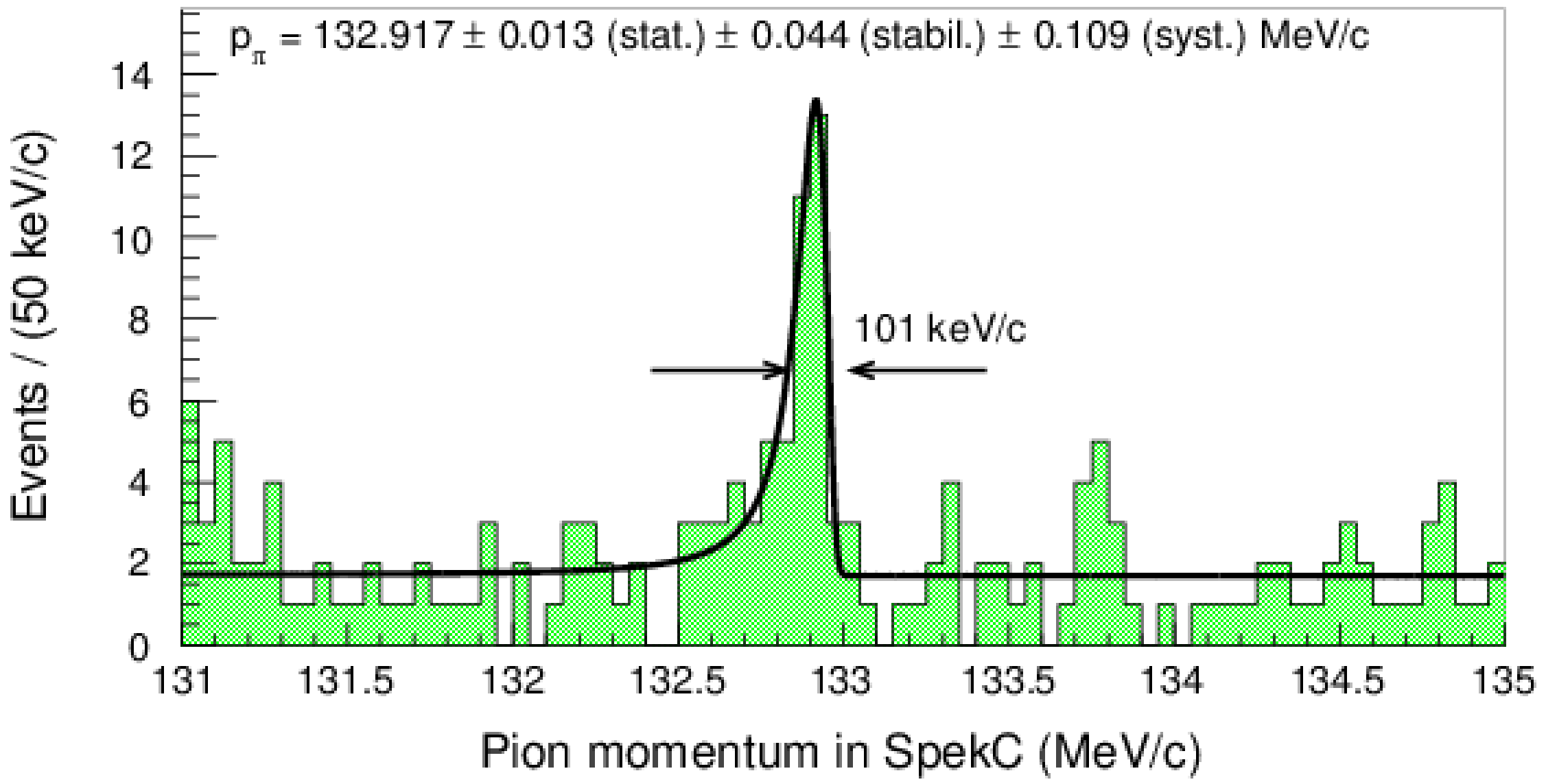

FIG. 2. Pion momentum distribution in SpekC for true coincidences with Kaos. A monochromatic peak at 133 MeV/c was observed which is a unique signature for the two-body decay of 50 stopped hyperhydrogen ${}^{4}_{\Lambda}H \rightarrow {}^{4}He + \pi^{-}$.

Statistical model predictions for the hyperfragmentaion yields from for ${}^{6,7}Li$ and ${}^{9}Be$ target nuclei, with subsequent two-body pionic decay at rest, are presented in Table 2 (for details see [13]).

Because of the large two-body branching ratio, ${}^{4}_{\Lambda}H$ is the most likely hyperfragment to be produced. All other hypernuclear decays are produced with significantly lower rates and, therefore, only ${}^{4}_{\Lambda}H$ could be detected with very low statistics in the first experiments. For observation of other hyperfragments, in particular the hypertriton ${}^{3}_{\Lambda}H$ or ${}^{6}_{\Lambda}H$, one needs to increase the luminosity of the experiment significantly. An increase of the luminosity by about an order of magnitude can be achieved by using a new target geometry at MAMI [13] (see FIG.

1 (right)) for a schematic view of the target and detection geometry). Instead of employing a tilted target, the beam would be incident on a 1.5mm wide surface of a 50 mm long target block, aligned along the beam axis. Pions emitted at 90 degrees with respect to the beam direction toward the spectrometers, would traverse only a thin layer of target material, keeping their momentum spread acceptably low.

Table 2: Statistical model predictions for the probabilities in % of forming a hyperfragment ${}^{A}_{L}Z$ with subsequent two-body $\pi^-$ decay [13].

| | | ${}^{9}Be$ target | | | ${}^{7}Li$ target | ${}^{6}Li$ target |
|---|---|---|---|---|---|---|
| ${}^{A}_{\Lambda}Z$ | $p_\pi$ (MeV/$c$) | ${}^{9}_{\Lambda}Li^*$ | ${}^{8}_{\Lambda}Li^*$ | ${}^{8}_{\Lambda}He^*$ | ${}^{7}_{\Lambda}He^*$ | ${}^{6}_{\Lambda}He^*$ |
| ${}^{3}_{\Lambda}H$ | 114.37 ± 0.08 | 0.56 | 1.18 | 0.67 | 1.49 | 2.48 |
| ${}^{4}_{\Lambda}H$ | 132.87 ± 0.06 | 3.56 | 3.74 | 7.51 | 12.61 | 8.81 |
| ${}^{6}_{\Lambda}H$ | 135.13 ± 1.52 | 0.03 | <0.01 | 0.23 | 0.10 | _ |
| ${}^{6}_{\Lambda}He$ | 108.47 ± 0.18 | 2.44 | 1.25 | 1.47 | 3.53 | _ |
| ${}^{7}_{\Lambda}He$ | 114.97 ± 0.15 ± 0.17 | 2.12 | 0.44 | 1.35 | _ | _ |
| ${}^{8}_{\Lambda}He$ | 116.50 ± 1.08 | 0.04 | _ | _ | _ | _ |
| ${}^{7}_{\Lambda}Li$ | 108.11 ± 0.05 | 1.54 | 1.68 | _ | _ | _ |
| ${}^{8}_{\Lambda}Li$ | 124.20 ± 0.05 | 0.85 | _ | _ | _ | _ |

## 3. DECAY PION SPECTROSCOPY OF THE Λ-HYPERNUCLEI: A NEW APPROACH

In the experiments described above (Fig.1), the produced kaons and decay pions were detected simultaneously by magnetic spectrometers Kaos for kaons ($HKS$) and SpekA, SpekC for pions ($H\pi S$).

The detection probability $R_{K^+}(HKS)$ of produced $K^+$ in the momentum acceptance $P_0 \pm \Delta P$ of $HKS$ is: $R_{K^+}(HKS) = [\Delta\Omega/(4\pi)] \times \varepsilon_s^K \times \varepsilon_{ef} \approx 10^{-3}$, were $\Delta\Omega = 20 - 30$ msr is the acceptance of $HKS$, $\varepsilon_{ef} \cong 0.5$ is the survival probability of the $K^+$, $\varepsilon_{ef} \cong 0.5$ is the detection efficiency of the detector package, and the accepted momentum bite $\Delta P \cong 100$ MeV/c. A similar figure is obtained for the detection probability of decayed pions: $R_{\pi^-}(H\pi S) \approx 10^{-3}$. The two events are correlated in time, and their simultaneous detection probability is $R_{K^+\&\pi^-} = R_{K^+}(HKS) \times R(\pi^-) \times R_{\pi^-}(H\pi S)$, where $R(\pi^-) \leq 10^{-2}$ is the probability of decay pion formation for each produced $K^+$ [8]. Therefore, for this type of experiment the useful event probability for each produced $K^+$ is $R_{K^+\&\pi^-} \approx 10^{-8} / K^+$.

The useful event rate can be increased by several orders of magnitude by using the concept of "*delayed-$\pi^-$*" spectroscopy [14]. The experimental setup in this case consists only of the decay pion spectrometer, $H\pi S$, i.e. SpekA and SpekC in FIG. 1. The tracking detector package of $H\pi S$ is the same as before, but an ultra-high precision timing detector, based on Cherenkov radiators coupled to Radio Frequency Photomultiplier Tubes (RF PMT) [15] will be added. The expected time resolution of such detectors is better than 10 ps, FWHM. The expected reconstructed transit-time spread of pions in HπS is ~ 20 ps, FWHM, and consequently the expected total time resolution is also ~ 20 ps, FWHM. The time structure of the CW electron beams at Jefferson Lab and MAMI has beam buckets of ~ps duration separated at ~ns intervals. By correlating the decay pion times with the particular beam bucket that caused the event, they can be separated from the huge amount of promptly produced background, without detecting $K^+$ mesons as a positive tag of strangeness production. Indeed, the probability to find promptly produced (non-decay) pions at times larger than 100 ps is less than $10^{-5}$, while ~70% of decay pions from Λ or hyperfragments (lifetime 260 ps) are delayed more than 100 ps. Therefore, the $H\pi S$ equipped with RF PMT based Cherenkov detectors will make "delayed $\pi^-$spectroscopy", similar to "delayed γ-ray spectroscopy", a feasible technique. The useful, decay pion event probability in a "delayed $\pi^-$spectroscopy" experiment is: $R_{\pi^-} = 0.7 \times R_S \times R(\pi^-) \times R_{\pi^-}(H\pi S)$, were $R_S$ is the probability of strangeness production. In this case, all two body reactions, were the virtual photon produces strangeness, contribute. These reactions are: $1{:}\, \gamma + p \to \Lambda + K^+$; $2{:}\, \gamma + p \to \Sigma^0 + K^+$; $3{:}\, \gamma + p \to \Sigma^+ + K^0$; $4{:}\, \gamma + n \to \Lambda + K^0$; $5{:}\, \gamma + n \to \Sigma^- + K^+$; $6{:}\ \gamma + n \to \Sigma^0 + K^0$.

In addition the entire useful virtual photon spectrum participates, unlike previous experiments where only virtual photons producing $K^+$ with momentum inside the $HKS$ acceptance contributed. Due to this, the expected rate of useful events for "delayed $\pi^-$ spectroscopy", would be $10^3$ times higher than the rate expected in previous experiments.

The RF PMT-Cherenkov detector will also provide an absolute momentum calibration of the $H\pi S$ by measuring the time-of-flight (TOF) differences of pairs of particles with different masses [15]. Monte Carlo simulations demonstrated that the magnetic spectrometers at the MAMI electron-scattering facility can be calibrated absolutely with an accuracy $\delta p/p \leq 10^{-4}$, which will be crucial for high precision determination of hypernuclear masses [15]. In addition by measuring lifetimes of the low lying hypernuclear states and by applying the "tagged-weak $\pi^{-}$ method" [16], the electromagnetic excitation rates of such states with lifetimes down to $\cong 10^{-11}$s could be investigated, which is another avenue for hypernuclear studies. The experimental information on these characteristics of the ΛN interaction is essential to test and improve baryon-baryon interaction models, which are needed for a comprehensive understanding of the strong interaction between hyperons and high density nuclear matter [14].

Here, we propose a new experiment for decay pion spectroscopy of light hypernuclei, which would be suitable both for electron- and proton-beam facilities. Identification of produced hypernuclei is realized by detecting the low-energy recoils and using the recoil distance technique. The general layout of the experimental setup is shown in FIG. 3. The incident electron (or proton) beam hits the 2 mg/cm$^2$ target and produces a hyperfragment, which exits the target and decays after ~ 200 ps outside the target, resulting in a decay pion and recoil nucleus. The flight distance from the primary interaction point is estimated to be ~ 1 mm. Decay pions are detected in the high-resolution magnetic spectrometer - $H\pi S$ located at ~90 degrees. In this case, the hyperfragment decays outside of the target, and as a result the monochromatic spectrum of produced pions is broadened by kinematics. The recoil nucleus is detected by means of the recoil detector, which is located at the opposite side of the beam to the $H\pi S$. This is a low-pressure (LP) MWPC, which is very insensitive to gamma rays and minimum ionising particles and has good position resolution for the low energy recoiling particles. In this case the decay pion event probability is: $R_{\pi^-}(rd) = \varepsilon(rd) \times R_S \times R(\pi^-) \times R_{\pi^-}(H\pi S)$, where $\varepsilon(rd)$ is the detection efficiency for the recoiling particle after a recoil distance analysis has been performed. It is expected that $\varepsilon(rd) \geq 0.1$, so that the useful, decay pion event probability is: $R_{\pi^-}(rd) \geq 0.1 \times R_{\pi^-}$, which increases event rate by about two orders of magnitude, compared to previous experiments. In addition this technique can also be applied to proton induced experiments.

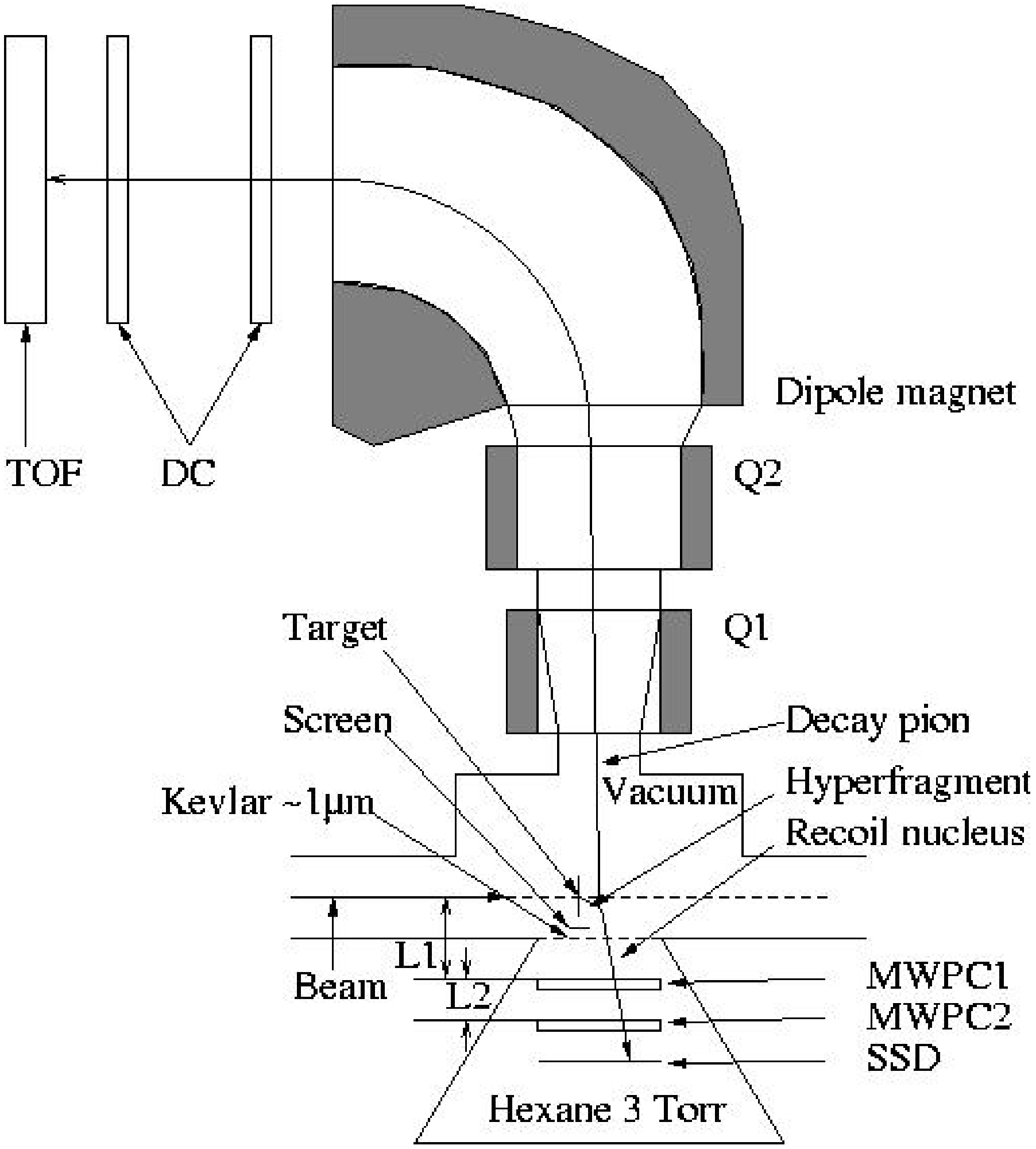

FIG 3. Schematic view of the proposed experiment consisting with the high-resolution pion spectrometer ( $H\pi S$) and the low-energy recoil detector for detection of the recoiling nuclei in coincidence with the decayed pions.

## 4. LOW-ENERGY RECOIL DETECTOR

A Recoil Detector (RD) is shown schematically in FIG. 3. It consists of two LP MWPC units (MWPC1, MWPC2) and a solid state detector (SSD) which form a position, time and energy sensitive low energy particle telescope. The LP MWPCs have an active area of 30×30 mm$^2$, a separation distance of L2 cm and are windowless to enable passage of low energy nuclear fragments, starting from heavy hydrogen isotopes. Heptane or Hexane at about 3 Torr pressure will be the MWPC gas. The distance from MWPC1 to the center of the experimental target is L1 cm and dimensions L1 and L2 (FIG. 3) will be determined in dedicated experimental studies. The target is a thin foil of thickness ~2mg/cm$^2$, so that most produced hyperfragments exit and decay outside of the target. The RD is screened from the central beam- on-target region, to enhance the signal for hyperfragments produced outside of the target, and separated from the accelerator vacuum by a 1-2 μm Mylar or Kevlar foil. In FIG. 3 the recoiling nuclei are detected by MWPC1, MWPC2 and the SSD, while decay $\pi^-$ mesons are detected by the magnetic spectrometer HπS. In principle the RD can be operated without an SSD. In that case the energy of the recoiling nucleus would be determined by its velocity with its type determined from the decay pion momentum.

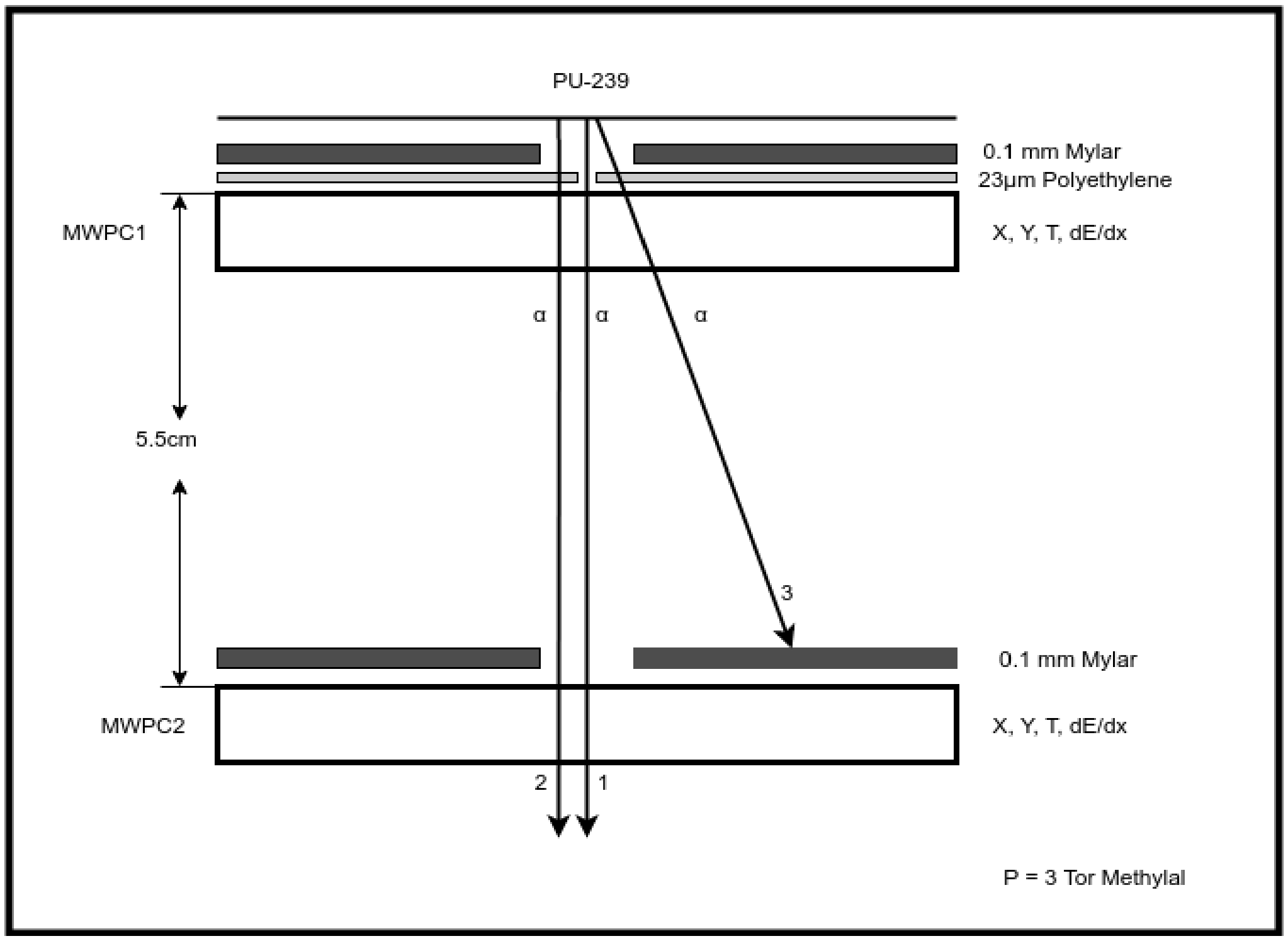

FIG. 4. Schematic representation of the test setup.

The architecture of the MWPCs and tests of a prototype detector using methylal ($(OCH_3)_2 CH_2$) or hexane ($C_6H_{14}$), at a pressure of a few Torr are presented in [17]. The test (FIG. 4) employed a 0.14 mg/cm$^2$ thick, 30×30 mm$^2$ $^{239}Pu$ α-particle source which has three lines: 11%-5.099 MeV, 20%-5.137 MeV and 69%-5.150 MeV. MWPC1 and MWPC2, separated by 5.5 cm, detected the α-particles, which were collimated by 1 mm holes bored in 100 μm mylar absorbers. Time, position and dE/dx distributions from the MWPCs were recorded. Data was also obtained for reduced α-particle energies, by placing a polyethylene absorber between the first Mylar collimator and the MWPC1. In this case the holes in the Mylar collimators had 7 mm diameter and the polyethylene absorber had a 3 mm hole, which allowed simultaneous detection of incident α-particles and α-particles after passing through the polyethylene absorber.

The low-pressure MWPC system can be operated in the so-called single-step and double-step modes ([17,18] and references therein). The traces of typical α-particles signals from the anode and cathode planes in double-step mode are displayed in FIG. 5.

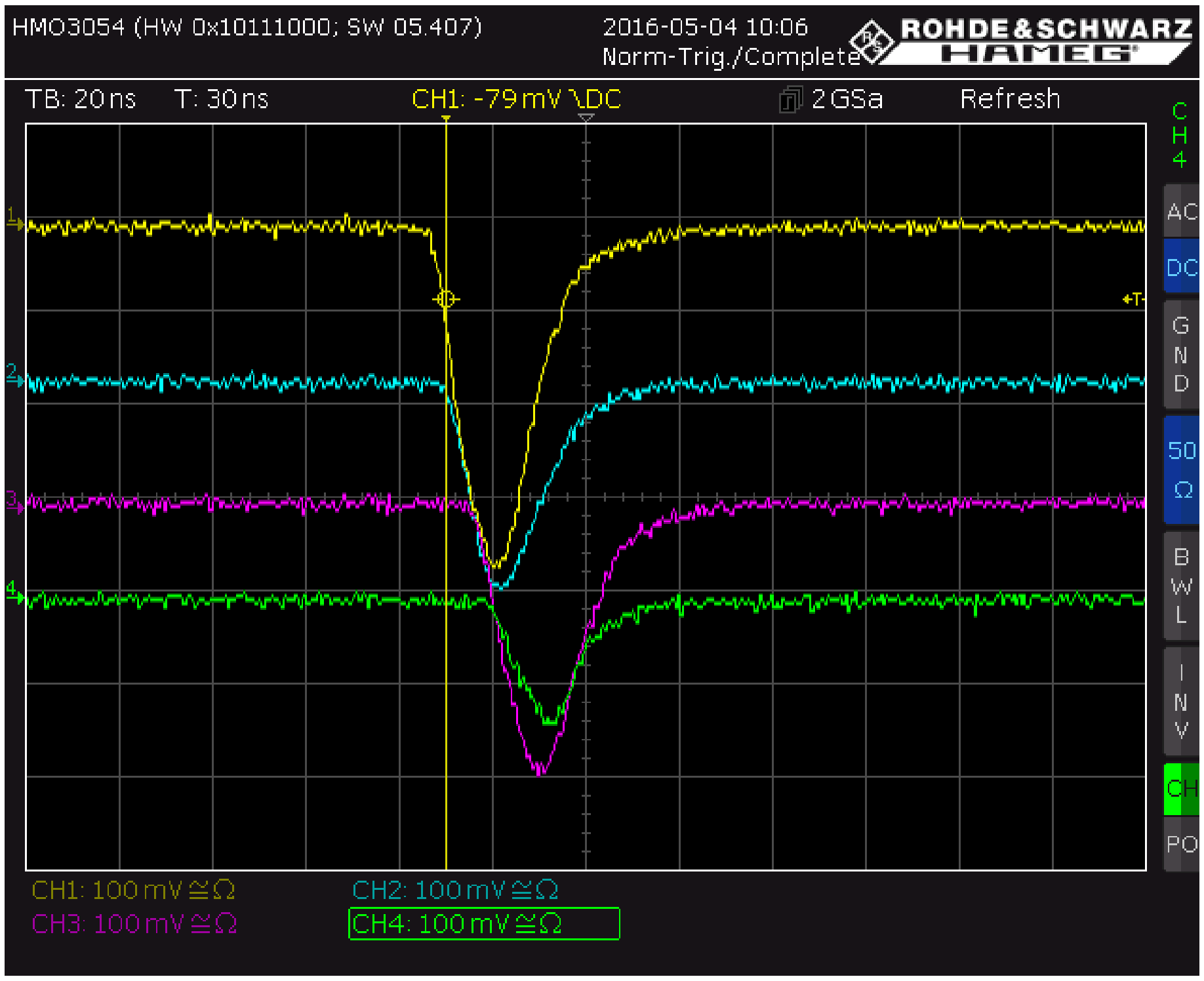

FIG. 5. Typical signals from anode and cathode planes after amplification, for operating conditions: hexane 2.6 Torr, double-step mode. Trace 1 (yellow), 2 (cyan): signals are from the anode planes of the top and bottom MWPC units. Trace 3 (magenta), 4 (green) signals are from the two cathode planes of the top MWPC unit. The timescale is 20 ns/division and amplitude scale 100 mV/division.

The time-difference distributions for signals from the anode planes of MWPC1 and MWPC2 are shown in FIG. 6 (without absorber) and FIG. 7 (with 23 μm polyethylene absorber). A Gaussian fit to the time-difference distribution of FIG. 6 has a standard deviation of 612 ps. Assuming that the timing resolutions of the two contributing MWPC units are equal, the time resolutions of the individual MWPC units are equal to $612/\sqrt{2} \approx 430$ ps.

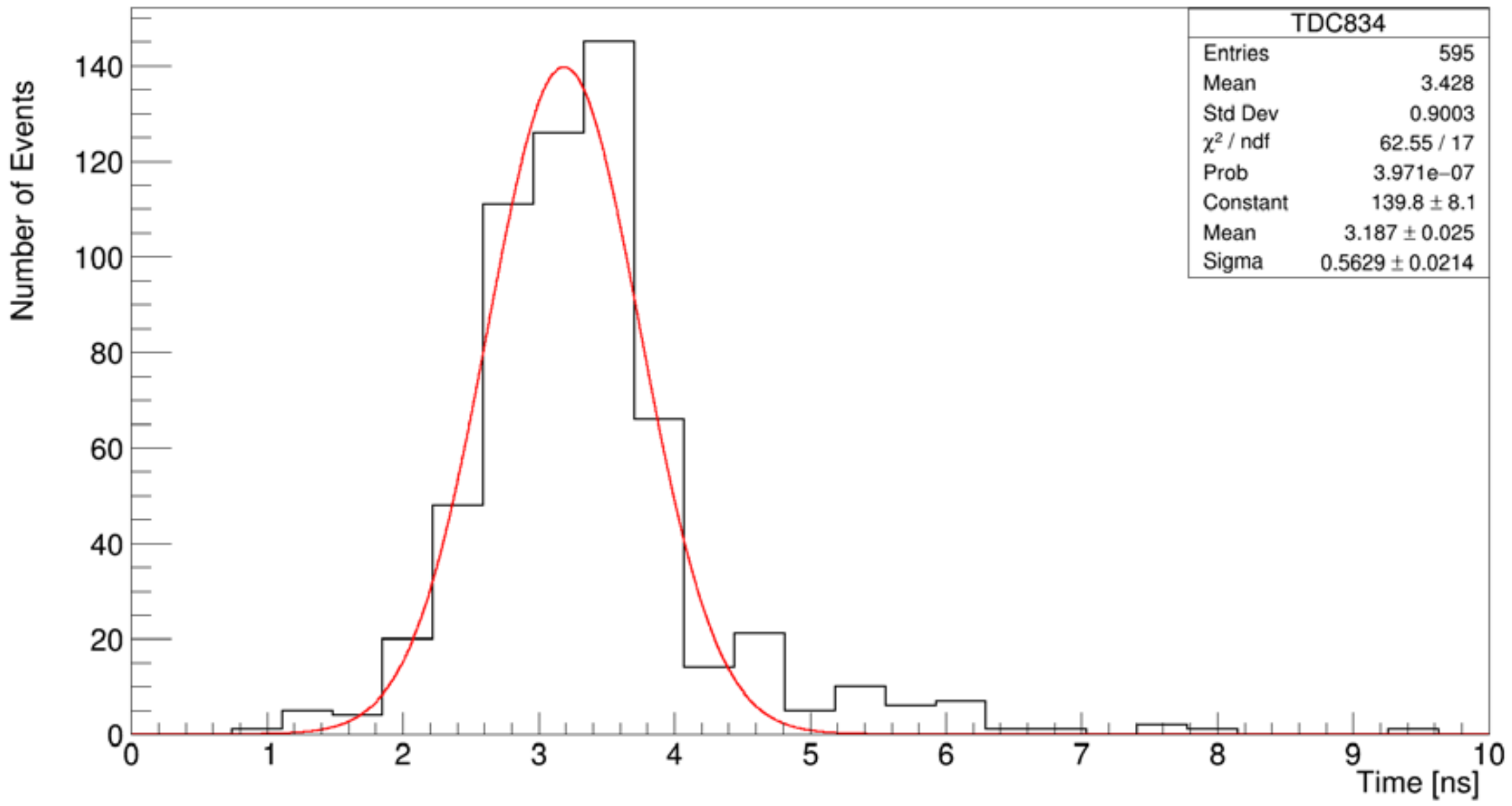

FIG. 6. Time difference between signals from the anode planes of MWPC1 and MWPC2.

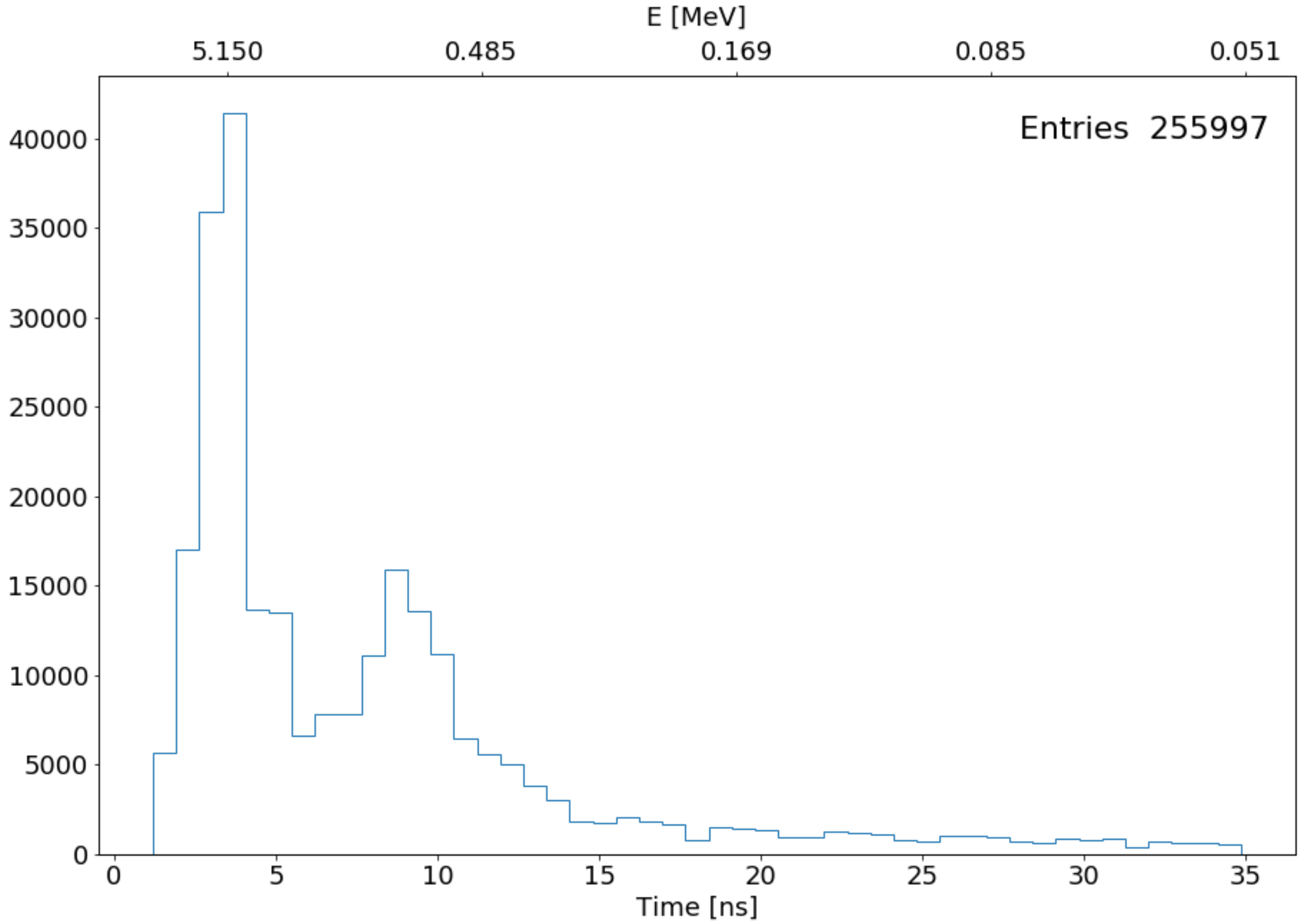

FIG. 7. As in FIG. 6 but with a 23 μm polyethylene absorber inserted. The top scale corresponds to the energies of α-particles.

In FIG. 8 the time-difference spectrum between signals from the anode and one of the two cathode lanes of MWPC1 is presented. The standard deviation of the Gaussian fit is about 1.2 ns, which results in 1.8 mm position resolution. With an alternative readout method [18], 100 μm precision can in principle be reached.

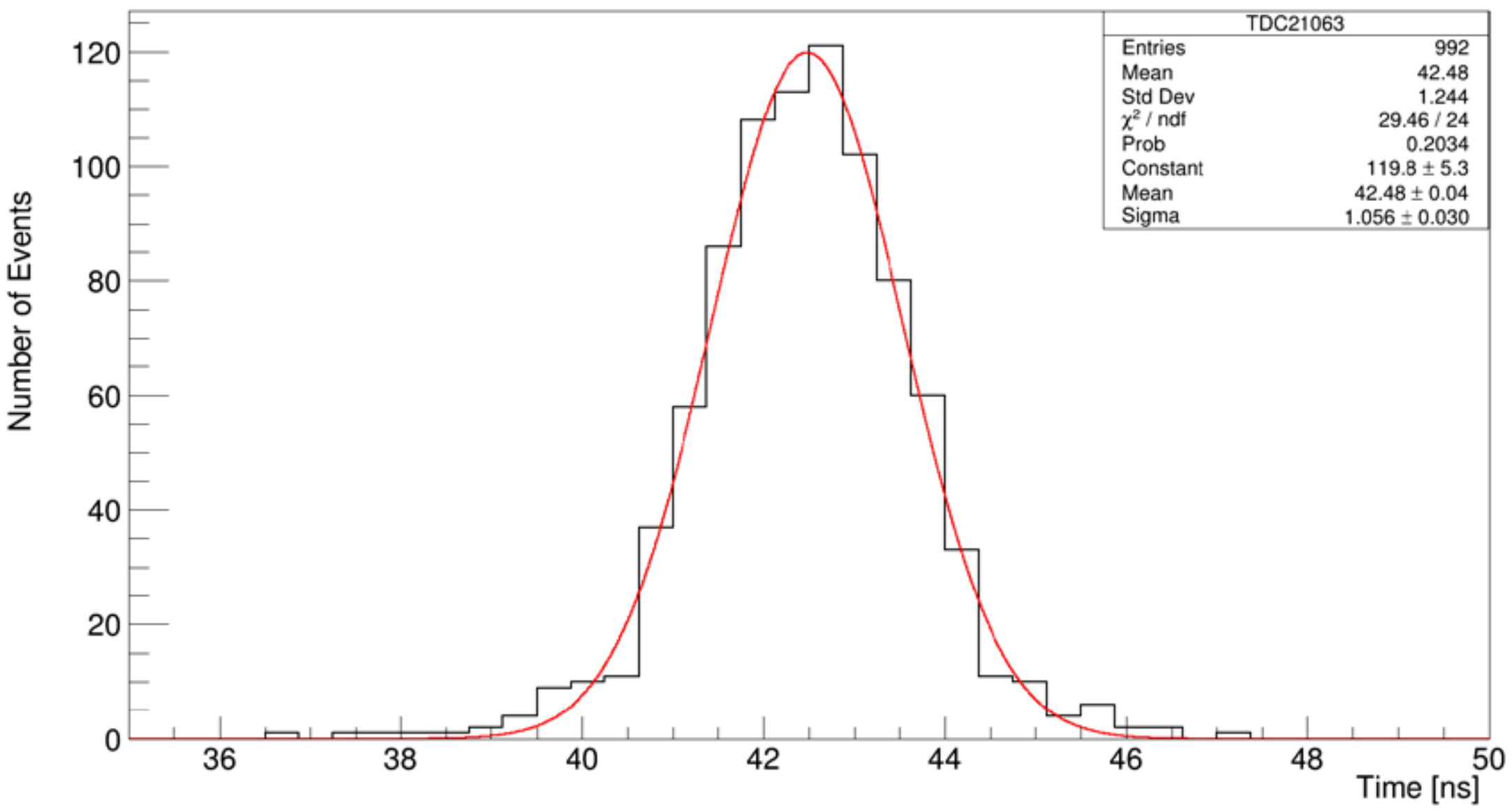

FIG. 8. Time difference between the anode and one of two cathode planes of MWPC1.

The time resolution of a MWPC unit in terms of the width of the time-difference distribution between the anodes of MWPC1 and MWPC2 was studied as a function of hexane pressure. This study demonstrated [17] that the best time resolution can be achieved at a pressure lying in the range of 2-3 Torr.

To be detected, low-energy recoils, i.e. low-energy $^{3}He$, $^{4}He$, $^{6}He$ (and heavier species) require enough energy to pass through the Kevlar window and both MWPC units. In Fig. 9 the stopping powers (differential energy losses) and ranges for $H, He$ and $C$ ions in hexane, simulated using the SRIM software package [19] are displayed. FIG.10 presents the corresponding distributions for Kevlar.

From FIG. 9 and 10 we calculate that a LP MWPC based RD system (FIG. 3) filled to 3 torr with hexane is capable of detecting $^{3}He$, $^{4}He$, $^{6}He$ recoils with energies above 35, 50, 70, keV and $^{12}C$ nuclei with energies higher than 100 keV.

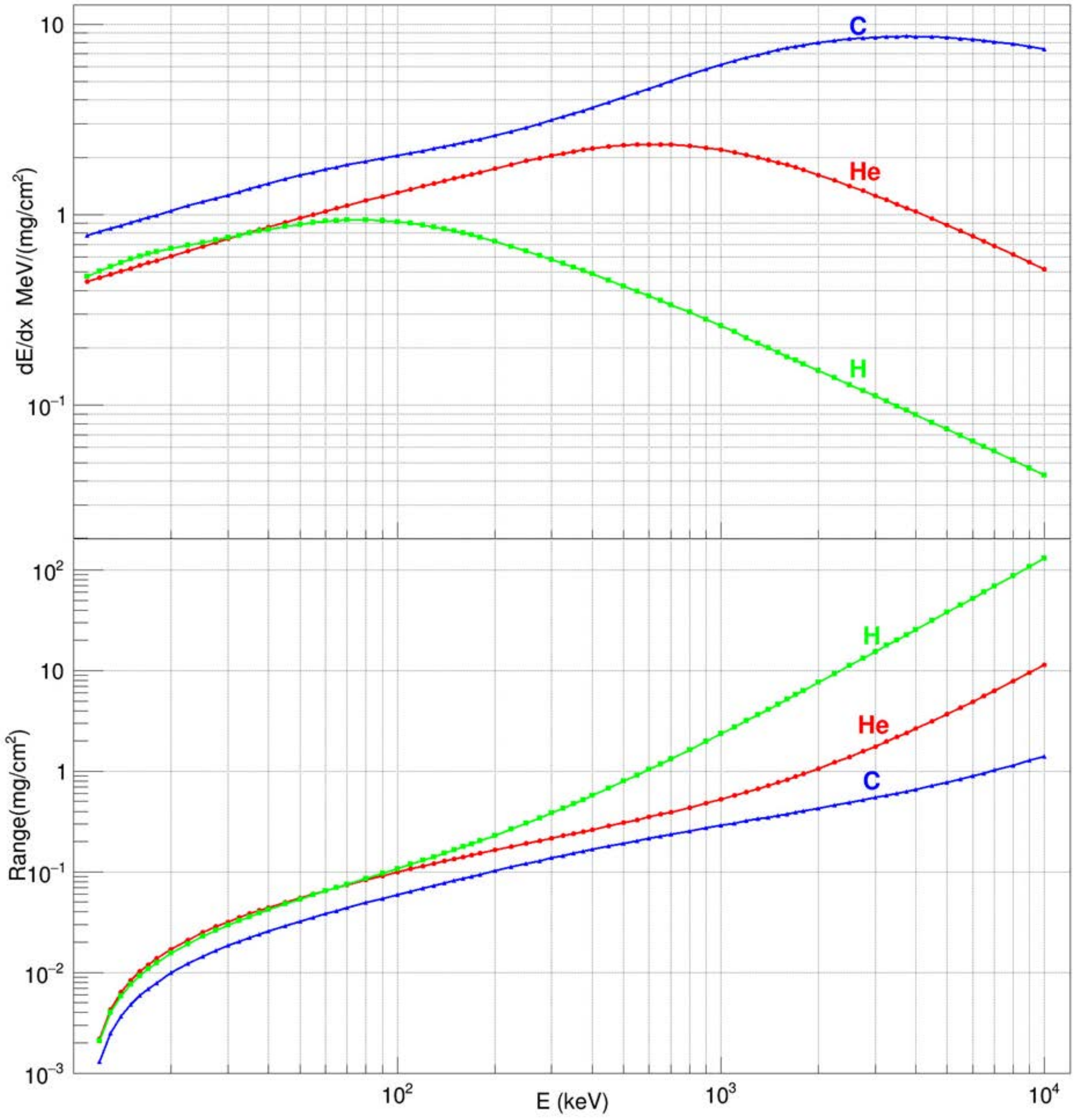

FIG. 9. The stopping powers and ranges of H, He and C ions in Hexane vs. energy.

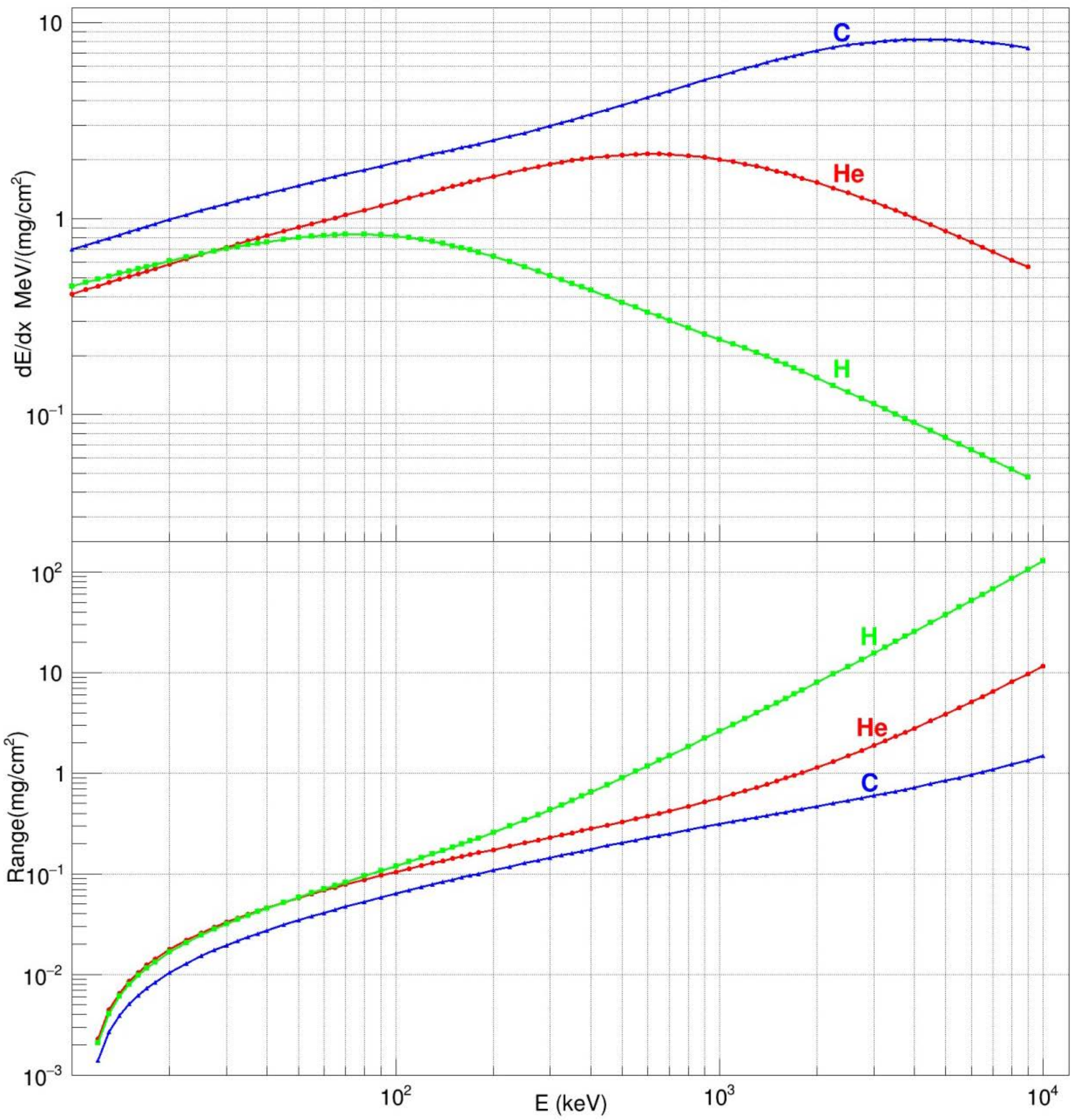

FIG. 10. The stopping powers and ranges of H, He and C ions in Kevlar vs. energy.

Since the RD will be located near to the incident beam to maximize the geometrical acceptance it is expected that many background particles, e.g. due to electromagnetic processes, will be produced. Even though the sensitivity to electrons and photons is very low, the rate capability of the MWPC units is a crucial factor. Typical signals from the anode plane of MWPC1, placed next to the Pu-239 source without any collimator, are displayed over a time scale of around 12 μs in FIG. 11. From the 21 signals registered cleanly during the 12 μs oscilloscope sweep, we infer that the MWPC has the capability to operate properly at rates in excess of 1 MHz.

Given that huge numbers of electron-positron pairs will be produced by beam interactions, the sensitivity of the low-pressure MWPC to minimum ionizing particles was measured by replacing the Pu-239 α -particle source with a $\beta$-particle source (Sr-90). The rejection factor for a single MWPC unit, i.e., the ratio of $\beta$-particle to α-particle detection efficiencies is about $10^{-4}$ for the 30 mV thresholds used typically in the electronics. Since the recoil nuclei will be detected in two MWPC units, the overall $\beta$-particle/α-particle detection efficiency will reduce to ~$10^{-8}$. A more quantitative value for this rejection factor and other characteristics of the RD will be obtained under realistic experimental conditions in future test experiments.

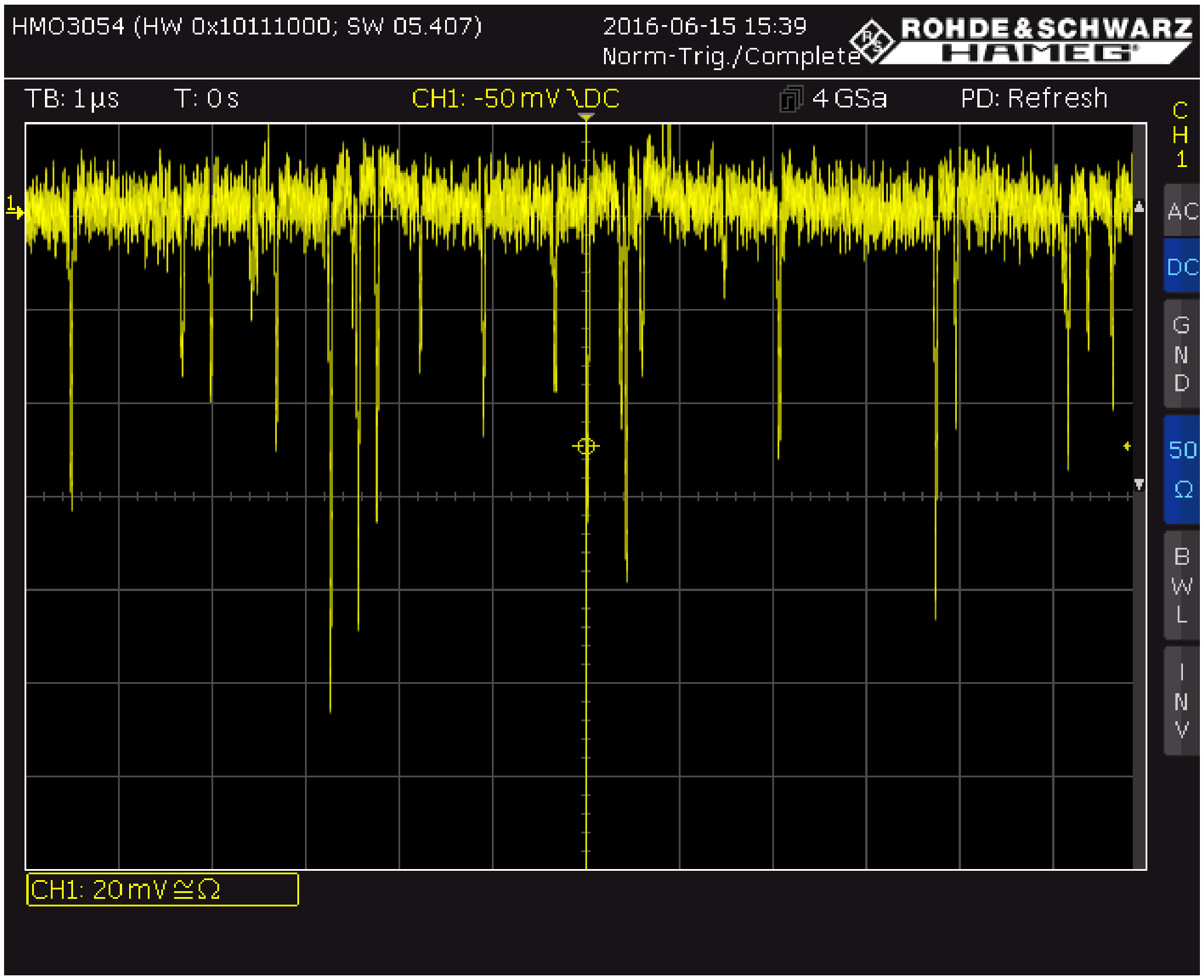

FIG. 11. Typical signals from the anode plane of MWPC1, displayed on a 500 MHz oscilloscope. The α-particle source is Pu-239, with no collimation applied. MWPC conditions are: hexane 2.85 Torr and double step operation.

## 5. MONTE CARLO SIMULATION STUDIES

The Monte Carlo simulation consists of a hypernuclear event generator which feeds particles into a model of the detection apparatus. Using the simulated response (energy, time, position etc.) of the detectors, the effective masses and binding energies have been reconstructed in a manner analogous to a real experiment. The hypernuclear mass ($M_{\mathrm{HY}}$) was calculated using the kinematic relation:

$$M_{\mathrm{HY}} = \sqrt{\left(m_R^2 + m_\pi^2 + 2 \times \sqrt{p_R^2 + m_R^2} \times \sqrt{p_\pi^2 + m_\pi^2} - p_R p_\pi \cos(\psi)\right)}, \qquad (3)$$

Where $m_R$ and $m_\pi, p_R$ and $p_\pi$ are the masses and momenta of the recoil nucleus and decay pion, respectively and $\psi$ is the angle between recoil nucleus and pion (the so called folding angle). Thus there are three kinematic variables to be measured: $p_\pi, p_R$ and $\psi$. In the Monte Carlo simulations we assumed the following:

1. The momentum distribution of the produced hypernuclei is assumed to be isotropic in angle with uniformly distributed magnitude in the range 0-100 MeV/c. Hypernuclei are assumed to be produced outside of the target.
2. The relative momentum resolution ($\Delta p/p$) of the pion spectrometer ($H\pi S$) is described by a Gaussian of width $\sigma_p = 10^{-3}$.
3. The energy resolution of the RD is described by a Gaussian of width $\sigma_E = 50$ keV.
4. The folding angle precision is described by a Gaussian of width $\sigma_\psi = 2$ degree.

A simulated distribution of the folding angle between $^3He$ and $\pi^-$ for ${}^3_\Lambda H \to {}^3He + \pi^-$ decays is shown in FIG. 12. This distribution shows that the $^3He$ recoil nuclei are directed in the angular region around 180 deg. relative to the $\pi^-$ direction. This can be used to optimize the design of the geometry of the RD.

Simulated distributions of recoil-nucleus energies, decay-pion momenta, binding energies and track lengths of hypernuclei for ${}^3_\Lambda H \to {}^3He + \pi^-$ and ${}^4_\Lambda H \to {}^4He + \pi^-$ decays are shown in FIG. 13 and FIG. 14 respectively. For the ${}^6_\Lambda H \to {}^6He + \pi^-$ decays we assume the $\Lambda$ binding energy for ${}^6_\Lambda H$ to be 4.18 MeV which is the $\Lambda$ binding energy of the ${}^6_\Lambda He$ hypernucleus. The resulting distributions are shown in FIG. 15.

The resulting resolutions ($\Delta p/p$) for decay-pion momenta arise from two body kinematics and lie in the range 1-2% (r.m.s.). The resolutions for binding energies, which are calculated on the basis of Eq. 3, are about 120 keV (r.m.s.).

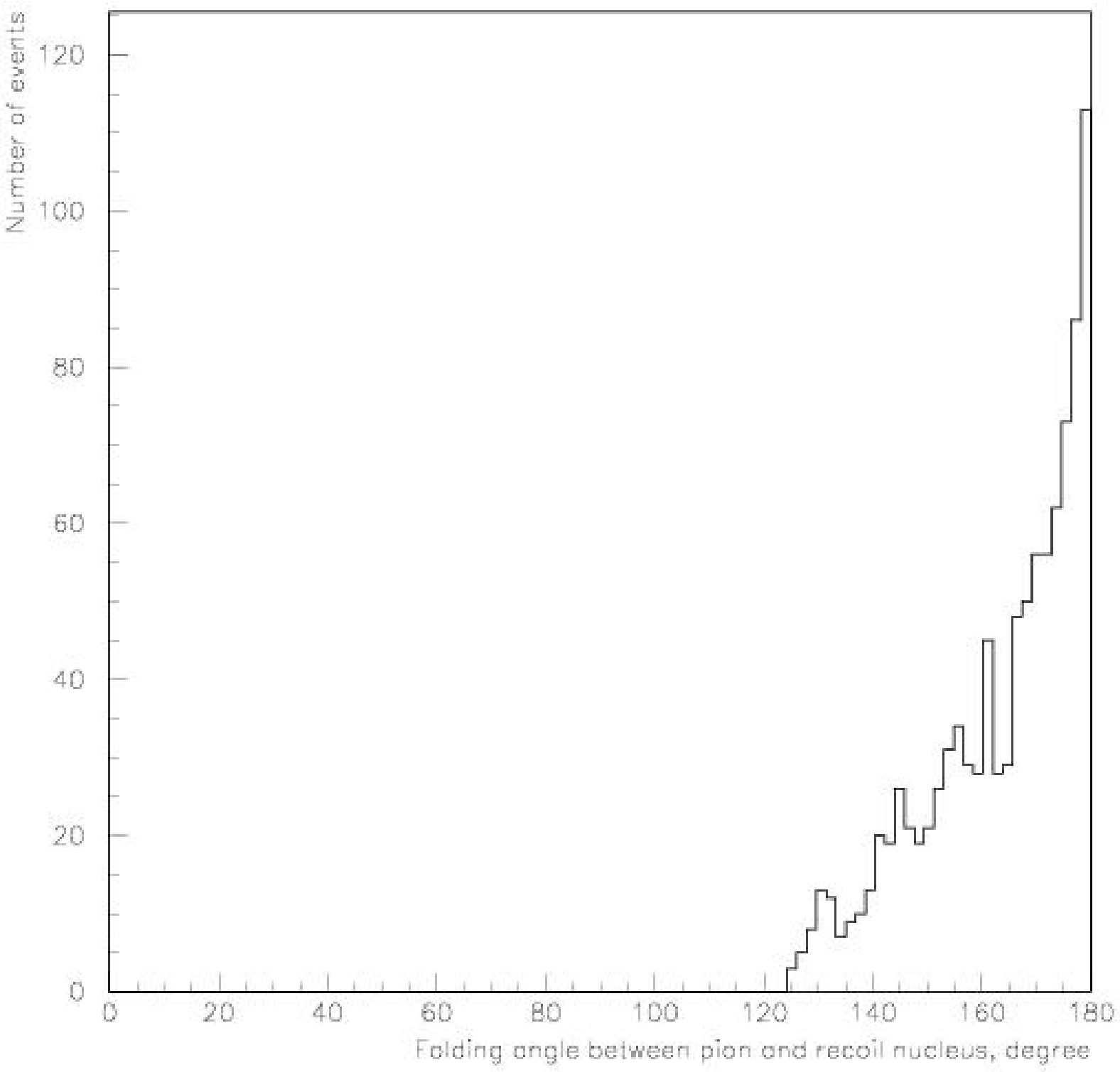

FIG. 12. The simulated folding angle distribution between the pion and recoil nucleus, for ${}^{3}_{\Lambda}H \rightarrow {}^{3}He + \pi^{-}$ decays.

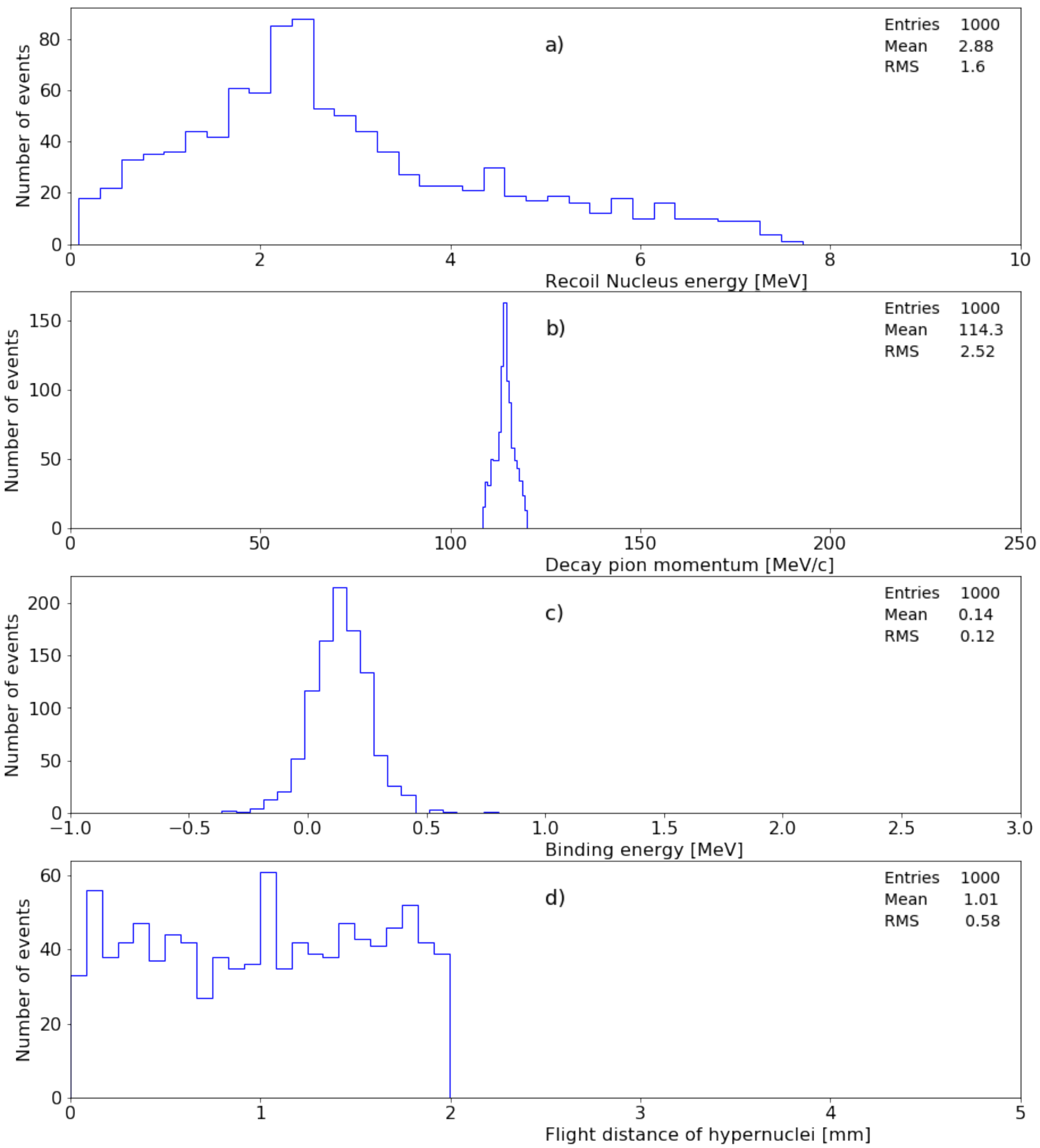

FIG. 13. Simulated distributions for ${}^{3}_{\Lambda}H \rightarrow {}^{3}He + \pi^{-}$ decays: energy of ${}^{3}He$ (a), decay pion momentum (b), binding energy (c), and flight distance of ${}^{3}_{\Lambda}H$ hypernuclide (d).

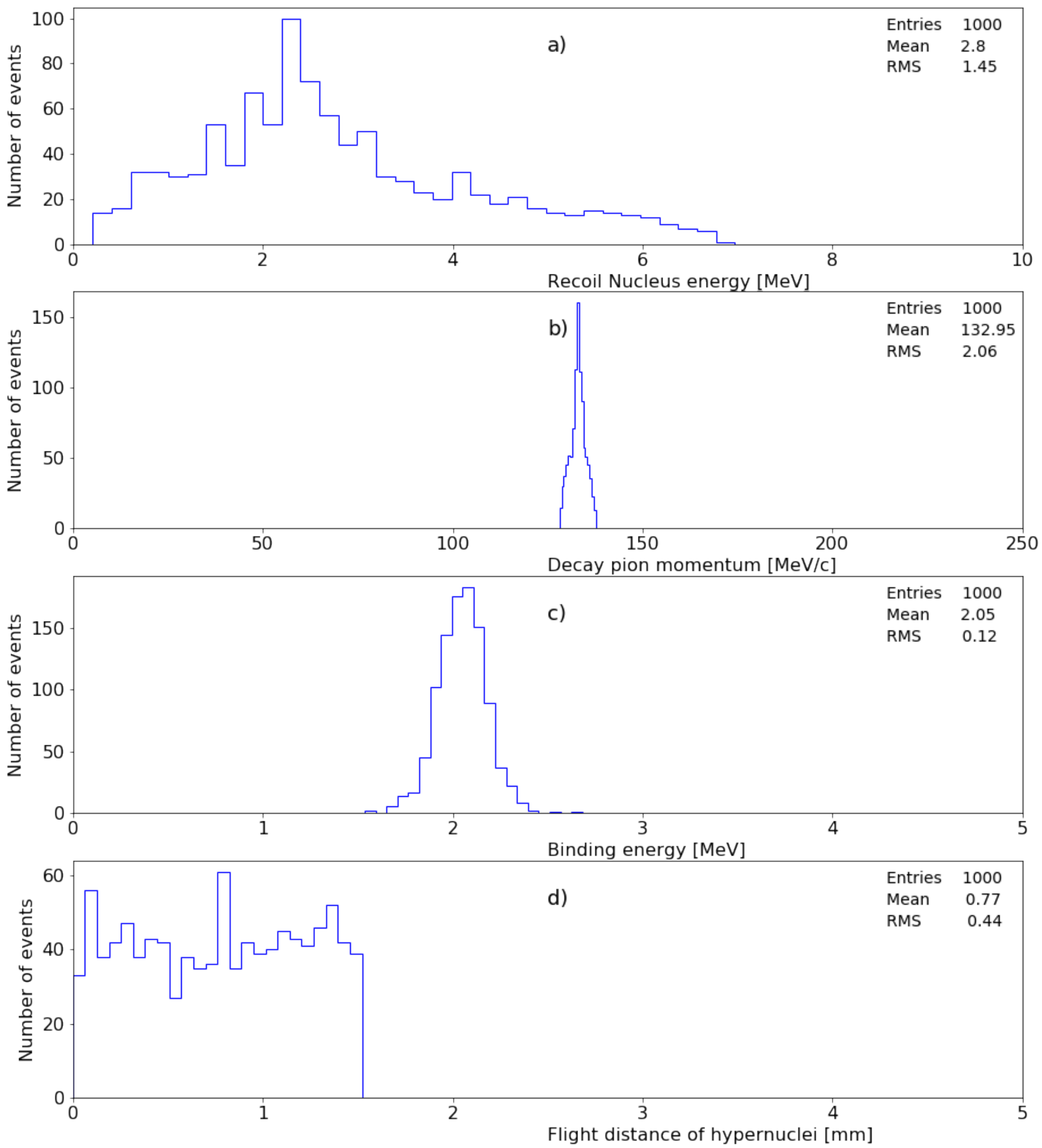

FIG. 14. Simulated distributions for ${}^{4}_{\Lambda}H \rightarrow {}^{4}He + \pi^{-}$ decays: energy of ${}^{4}He$ (a), decay pion momentum (b), binding energy (c), and flight distance of ${}^{4}_{\Lambda}H$ hypernuclide (d).

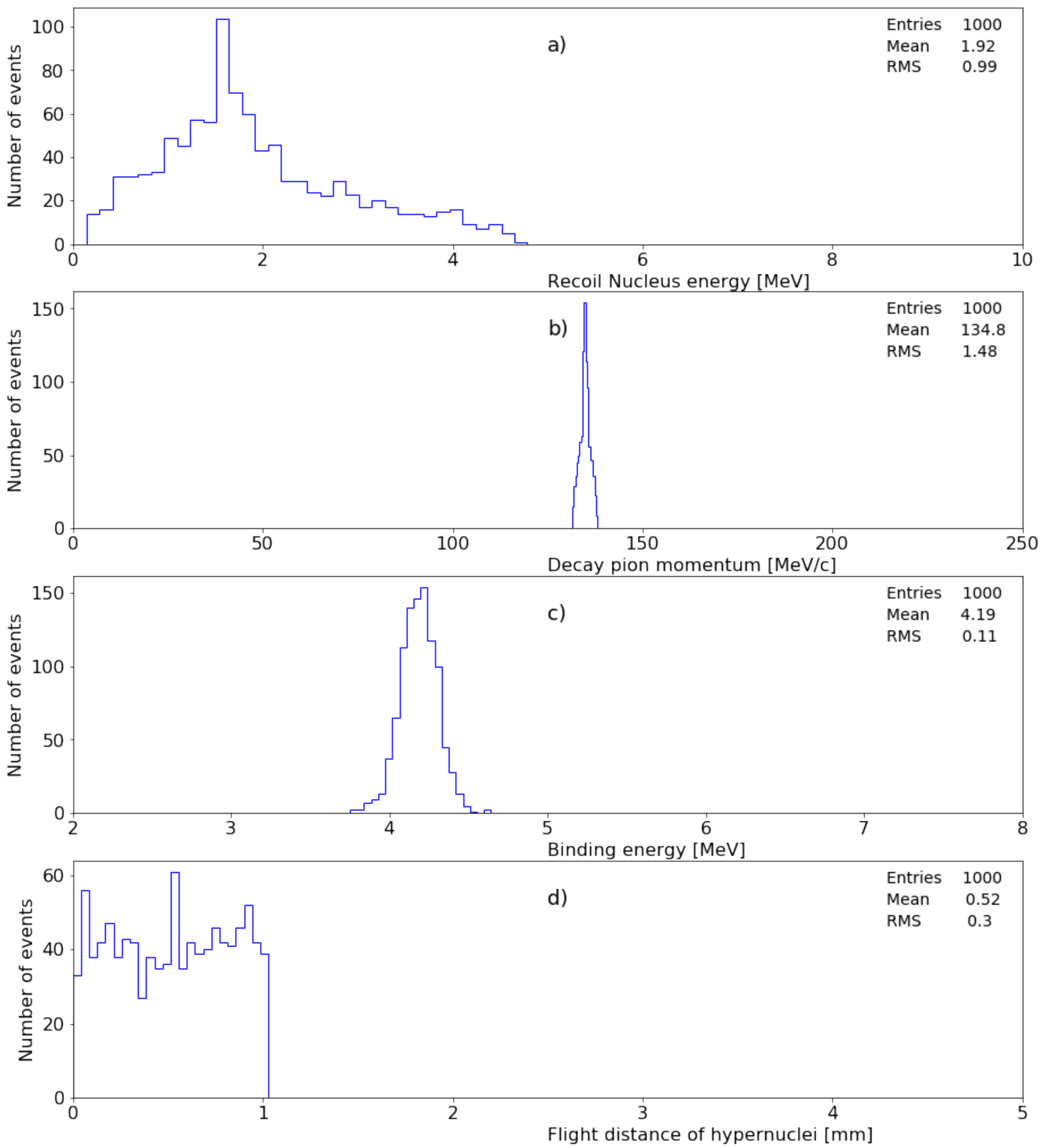

FIG. 15. Simulated distributions for ${}^{6}_{\Lambda}H \rightarrow {}^{6}He + \pi^{-}$ decays: energy of ${}^{6}He$ (a), decay pion momentum (b), binding energy (c), and flight distance of ${}^{6}_{\Lambda}H$ hypernuclide (d).

## 6. BACKGROUND

In real experimental conditions the monochromatic pion spectra will be masked by a huge amount of background originating from three sources:

a) Promptly produced pions;

b) Decay pions from $\Lambda$ particles produced by "quasi-free" processes;

c) Accidentals.

The promptly produced pions will be excluded by detection of the recoil nuclei, using a "vertex cut" which requires the reaction point (in the target) and the decay point (outside of the target) to be separated. The RD, which will be screened from direct view of the target, will provide precise determination of the trajectories of recoiling nuclei.

The "quasi-free" background will be suppressed by identification of the heavy recoiling particle species by the RD.

After screening from the target, the RD rate will be much lower than that of the HKS (Sec.2) which was used in previous Mainz experiments to identify hypernuclear events. As a result the level of accidentals between the RD and HπS will be much lower than was observed previously between the HKS and HπS.

Detailed MC studies of background processes, which include realistic models of experimental geometry and the RD response, are in progress.

## 7. YIELD ESTIMATION

### 7.1 Yield estimate for experimental studies at MAMI

In the first experiment at MAMI, a 1.508 GeV, 20 μA electron beam was directed at a 125 μm thick $^{9}$Be target, tilted at 54° to give an effective thickness of 27.7 mg/cm2. About 50 events from the two-body decays of stopped hyperhydrogen $^{4}_{\Lambda}\mathrm{H} \rightarrow {}^{4}\mathrm{He} + \pi^{-}$ were observed in a period of ~150 h. On this basis we estimate the rates for the proposed new experiment. The target in a new experiment must be about 10 times thinner to allow hyperfragments to exit and decay outside of the target. However the decay pion detection probability in similar experimental conditions (Sec. 2) is more than two orders of magnitude higher than previous experiments. Therefore the event rate will be increased more than order of magnitude. In addition the quasi-free background will be eliminated, since recoil detection excludes weak decay pions from quasi-free produced Λ-particles.

### 7.2 Yield estimate for internal target experiments

Electron-beam experiments would also be possible at an internal-target facility on a stored-beam accelerator, such as the Yerevan electron synchrotron. Some characteristics of the Yerevan electron synchrotron are listed in Table 3 and are used to estimate potential hyperfragment yields at this facility.

Table 3: Parameters of the Yerevan electron synchrotron.

| | |
|---|---|
| Energy (GeV) | 1.5 |
| Mean radius (m) | 34.49 |
| Mean orbit length (m) | 216.72 |
| RF frequency (MHz) | 132.79 |
| Circulating current (mA) | 22 |
| Intensity (electron/cycle) | $10^{11}$ |
| Intensity (electron/second) | $5 \times 10^{12}$ |
| Number of accelerated cycles Injection frequency? | 50 |
| Revolution period (μs) | 0.723 |
| Electron bunch length (ns) | 0.7 |
| Number of bunches | 96 |
| Electron internal target interaction time (ms) | 2 |
| Duty factor (%) | 10 |

We will consider a 1.5 GeV electron beam and a $^{9}Be$ target with thickness of 2.8 mg/cm$^2$. For stored-beam electron accelerators, the effective target length is equal to about 0.1 radiation length (r. l.) of the employed target material, due to the multiple passing of electrons through the target. This means that the effective length of the thin $^{9}Be$ target becomes ~ 6.5 g/cm$^2$ equivalent to around 2300 beam passes through the target. For a modest 3 mA stored electron beam a 2.8 mg/cm$^2$ thick target (which is equivalent to $7 \times 10^{11}$ electron/second and 6.5 g/cm$^2$ target thickness) and with a similar experimental setup to that described in Sec.3 for the extracted beam situation, we estimate 500(events)×6.5(g/cm$^2$)×0.11(μA)/20(μA)/.0028(g/cm$^2$) ≈ 6000 events in a period of 150 hr, i.e. more than an order of magnitude increase in event rate compared to an extracted-beam experiment. This kind of experiment could also be realized at the stretcher-booster ring of Tohoku University [20].

It is also worth mentioning that the first experiments performed at the Yerevan electron synchrotron were small-angle elastic electron scattering from the proton and deuteron, where the recoiling proton or deuteron was detected using solid state detectors [21].

### 7.3 Yield estimation for experimental studies at proton beams

The proposed experiment can also be realized with proton beams. In this case the expected hyperfragment production rates can be estimated from old emulsion data [22] or from recent theoretical predictions [23]. In Ref. [23] the theoretical production cross section of a $^{12}C$ beam impinging on a hydrogen target with energies from 1 to 10 GeV/nucleon is calculated for ${}^{2}_{\Lambda}n$, ${}^{3}_{\Lambda}n$, ${}^{3}_{\Lambda}H$, ${}^{4}_{\Lambda}H$, ${}^{6}_{\Lambda}H$, ${}^{7}_{\Lambda}Li$, and ${}^{7}_{\Lambda}Be$ hypernuclei and predicts that the ${}^{4}_{\Lambda}H$ production cross section in $^{12}C + {}^{1}H$ collisions at 5 GeV is about 10 μb. The two-body decay probability of heavy hyperhydrogen ${}^{4}_{\Lambda}H \rightarrow {}^{4}He + \pi^-$ is 0.5. Taking this into account the ${}^{4}_{\Lambda}H$ production and two body decay rate for a 5 GeV, 1 μA proton beam incident on a 2.8 mg/cm$^2$ thick $^{12}C$ target can be estimated to be $6\times10^{12}\times0.5\times10^{-29}\times2.8\times10^{-3}/12\times6.02\times10^{23}$/s = $4.34\times10^{3}$/s. For the detected decay pion rate we have $R_{\pi^-}(pC) = \varepsilon(rd) \times \varepsilon(\pi^-) \times 4.34 \times 10^3 \geq 0.4\, event/s,$ where $\varepsilon(rd) \geq 0.1$ and $\varepsilon(H\pi S) \sim 10^{-3}$ are the efficiencies of the RD and pion spectrometer with geometrical acceptances included. Even a modest 1 μA proton beam current produces a huge increase in the detected ${}^{4}_{\Lambda}H$ hyperfragment rate, namely 216000 in a 150 h beam time, compared to 50 events observed at MAMI with a 20 μA electron beam. Thus experiments with a proton beam will be ideally suited for the detection of the very rare hyperfragments such as ${}^{6}_{\Lambda}H$ and ${}^{8}_{\Lambda}H$. Recently ${}^{2}_{\Lambda}n$ and ${}^{3}_{\Lambda}n$ exotic hypernuclei were extensively discussed and looked for in relativistic ion experiments and also in an electron scattering experiment [24, 25, 26]. If such bound systems exist in nature they could be observed after a proton-carbon interaction by the mesonic decay processes ${}^{2}_{\Lambda}n \rightarrow d + \pi^-$ and ${}^{3}_{\Lambda}n \rightarrow t + \pi^-$ .

## 8. DISCUSSION

A new experiment for two-body $\pi^-$ decay spectroscopy of light hypernuclei is proposed. It employs a magnetic spectrometer for precise measurement of the decay pion momentum in conjunction with a detector of the recoiling nucleus for separation of produced hypernuclei from background. Low-pressure MWPCs are proposed as the position and time sensitive devices for low-energy recoil detection and their position and timing characteristics have been studied using

~5 MeV α-particles. The desirable features of this recoil detector include: good timing and position resolution; high rate capability; identification of heavy recoils; large acceptance ≥0.1; insensitivity to electromagnetic background and minimum ionizing particles.

Experiments of this type can be carried out using electron, photon or proton beams and $Li$, $Be$, or $C$ targets. Produced hyperfragments such as ${}^{3}_{\Lambda}H$ and ${}^{4}_{\Lambda}H$, ${}^{6}_{\Lambda}H$ and ${}^{8}_{\Lambda}H$ exit the target and decay in the vacuum. The monochromatic momentum spectrum of the produced pions is broadened by kinematics, but the absolute value is not changed by any secondary process such as ionization energy losses before detection in the magnetic spectrometer. Therefore average values of the $\pi^{-}$ momenta will reproduce exactly the two-body decay kinematics, so that they can be used for accurate determination of the binding energies of Λ-particles in light hypernuclei. This is especially crucial for the lightest hypernucleus, ${}^{3}_{\Lambda}H$, for which the binding energy of the Λ-particle is about 0.130 MeV.

For experiments with electron beams a simple analysis estimates that the expected yields of detected hyperfragments will be one-to-two orders of magnitude higher than was observed in previous experiments carried out at MAMI. Two-body $\pi^{-}$ decay spectroscopy of ${}^{3}_{\Lambda}H$ and ${}^{4}_{\Lambda}H$ hyperfragments using an RF structured electron beam and RF PMT based time-of-flight detectors will provide ultra-precise values of the binding energies of Λ-particles to ${}^{3}_{\Lambda}H$ and ${}^{4}_{\Lambda}H$ hyperfragments. These can be used to check theoretical predictions and as a reference for other hyperfragments.

Experiments with proton beams will have a very high yield compared to electron experiments. They can be used to extend the study of light hypernuclei to the exotic ${}^{2}_{\Lambda}n$ and ${}^{3}_{\Lambda}n$ hypernuclei and to the heavy hyperhydrogens ${}^{6}_{\Lambda}H$ and ${}^{8}_{\Lambda}H$.

In all cases, for realistic estimation of expected hyperfragment yields, test studies are needed to determine the maximum intensities of the incident electron/photon or proton beams at which the recoil detector will operate smoothly.

The described detector system may also be useful to study exclusive breakup reactions in low energy electron scattering experiments (e.g. 3He(e,e'd)p or 7Li(e,e'd)5He) where a low momentum recoil nucleus needs to be detected (see e.g. [27]).

## ACKNOWLEGEMENT

This work was supported by the RA MES State Committee of Science, in the frames of the research projects 14CYC-1c11, 16-A1c66 and 15T-2B206, the UK Science and Technology Facilities Council grant ST/L005719/1, the DAAD PPP, Germany 57345295/JSPS Research Cooperative Program, the Deutsche Forschungsgemeinschaft (DFG) Germany through Research Grant PO 256/7-1, and by the Grant-in-Aid for Scientific Research on Innovative Areas ”Toward new frontiers : Encounter and synergy of state-of-the-art astronomical detectors and exotic quantum beams”, JSPS/MEXT KAKENHI grant numbers JP18H05459 and JP17H01121.